\documentclass[english,twocolumn,prl,showpacs,linenumbers,floatfix]{revtex4-1}
\usepackage[latin1]{inputenc}
\usepackage{amsmath}
\usepackage{graphicx}
\usepackage{amssymb}
\usepackage{graphicx,color}
\usepackage{epsfig}
\usepackage{dcolumn}
\usepackage{bm}
\usepackage{lineno}
\usepackage{mathrsfs}
\usepackage{flushend}
\usepackage{cuted}
\newcommand{\nspace}{\!\!\!\!\!\!\!\!}
\usepackage{babel}

\begin{document}

\title{Theory of ac-Stark splitting in core-resonant Auger decay under strong x-ray fields}

\author{L. A. A. Nikolopoulos,  T. J. Kelly and J.  T. Costello}
\affiliation{School of Physical Sciences, Dublin City University and NCPTL, Dublin 9, Ireland}

\begin{abstract} 
In this work we report the modification of the normal Auger line shape under the action of an intense x-ray radiation. 
Under strong Rabi-type coupling of the core, the Auger line profile develops into a doublet structure with 
an energy separation mainly determined by the relative strength of the Rabi coupling. In addition, we find that the charge 
resolved ion yields can be controlled by judicious choice of the x-ray frequency.
\end{abstract}
\maketitle

\section{Introduction}
The interaction of an atomic system with a radiation field in the regime of x-rays  will lead to its ionization. 
The most dominant process will be, first, the ejection of  an inner-shell electron (photo-electron) with the absorption 
of a photon followed either by an intra-atomic Auger and/or a fluorescence transition. 
For relatively light atomic systems, the dominant decay channel of the single-hole singly-charged system is through an (radiationless) Auger 
transition, designated as 'normal Auger' which is a manifestation of electron-electron interaction.
One variation on this scenario is to  promote an inner-shell electron to an excited bound state, 
often denoted as Resonant Auger State (RAS), which can decay either through an Auger transition or by the emission of an x-ray photon. 
This process was first reported by Brown \cite{brown:1980} and since then a large number of investigations 
have taken place (see for example \cite{Armen:2000} and references there in).

Under excitation by the strong radiation fields, now available, from Free Electron Laser (FEL) sources 
\cite{wabnitz:2002,moller:2005,ackermann:2007,meyer:2010,young:2010}, the situation becomes considerably different.
Relevant studies in the context of strong laser fields have been reported quite early \cite{lambropoulos:1981} and 
in response to recent developments in the x-ray wavelength regime a number of theoretical and experimental works have already appeared
\cite{santra:2008, buth:2009,liu:2010,young:2010,cederbaum:2011}.
In the simplest situation, Rohringer and Santra in Ref. \cite{santra:2008} have studied the single-photon excitation of a neon 
K-shell electron to a RAS by an x-ray field and  a multipeak Auger Electron Spectrum (AES) is obtained for the fields 
they considered.
\begin{figure}[t]
\includegraphics[scale=0.35]{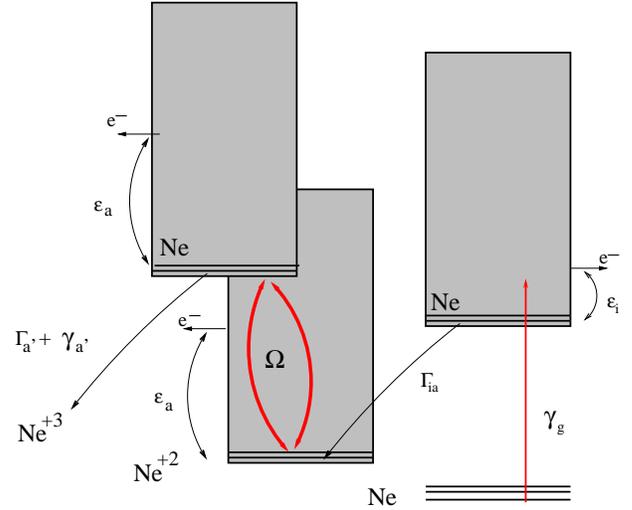}
\caption{(Color online) Schematic figure of the dominant channels involved in the interaction of neon with an x-ray field of 
frequency ca. 908.06 eV, chosen to match the Ne$^{+2}(1s^{-1}-3p)$ transition energy.}
\label{fig:fig1}
\end{figure}
In this work, we examine the AES and the ionic yields in the case where a normal Auger process takes place (as opposed to the RAS process) 
which involves photoionization of the neutral from the K-shell, followed by an Auger decay of the singly charged hole-system 
to the doubly charged ion. We demonstrate the emergence of ac-Stark splitting 
(also known as Autler-Townes splitting \cite{autler-townes:1955}) of the Auger resonance, resulting from strong Rabi-coupling 
of the apposite states in the resulting Ne$^{+2}$ ion. 
In contrast to RAS line-shape modification \cite{santra:2008}, the unusual phenomenon here is that the ac-Stark splitting 
is manifested in the kinetic spectrum of the Auger electron while strong Rabi-coupling occurs to one of the K-shell electrons 
of the doubly-charged ion. 
This effect requires an explanation on the basis of a two-electron representation of the Auger-electron ejection, 
instead of an ambiguous one-electron picture implied by the Rabi-type oscillation of an inner-electron. A detailed study 
of the two-electron representation, in a different context and formalism but similar physical background, 
can be found in Ref. \cite{hanson:1997}.

We have chosen as the target of our study neutral neon in it's ground state $| G \rangle = |$Ne$(1s^22s^22p^6,$ $^1S_0)\rangle$ and consider radiation 
with a photon energy of approximately $\omega = 908$ eV (see Fig. \ref{fig:fig1}). The ionization potential for the creation of a K-shell 
hole $| i \rangle = | $Ne$^{+}(1s2s^22p^6, $ $^1P)\rangle$ is $E^{(i)} = 870$ eV. Note that all the energies of the neon states are given relative to the 
neutral neon ground state.  Thus, the x-ray radiation will eject a K-shell photo-electron with a kinetic energy of ca. $\varepsilon_i = 38$ eV. Ionization from the outer shells is also possible but with much lower probability \cite{schmidt:1997}. The generated K-shell hole Ne$^{+}$ will 
decay, by filling the $1s$ vacancy, predominantly to the doubly charged neon state $| a \rangle $ = $|$Ne$^{+2}(1s^22s^22p^4,$ $^1D_2)\rangle$ 
with energy $E^{(a)} = 65.35$ eV, and Auger decay width $\Gamma_{ia} =  0.27$ eV \cite{schmidt:1997}. 
In addition, the Ne$^{+}$ ion with one K-shell hole can also decay through fluorescence by emitting x-ray radiation \cite{schmidt:1997}. 
This channel is about 55 times weaker than the Auger decay rate. Let's designate the sum of all decay channels of the $|i\rangle$ state as 
$\Gamma_i$. Normally, this Auger transition of the singly charged ion to the doubly charged ground state is accompanied by the ejection of 
an electron (Auger-elecron) with a kinetic energy $\varepsilon^{(0)}_a = E^{(i)}-E^{(a)} \sim 804.65$ eV and a Lorentzian line profile. 
In the present case we have chosen the photon energy to match the Ne$^{+2}(1s^{-1}-3p)$ transition energy.
Around this energy there is a manifold of excited states which we denote collectively as $| a' \rangle$. 
We show in table  1  those states which lie within a 1 eV band around the excitation photon energy. 
While the Auger state $|i\rangle$ decays with a lifetime of about $\Gamma_{ia}^{-1} \sim 2.44 $ fs, the field, 
through a Rabi-type transition, creates a coherent superposition of the ionic ground state $|a \rangle$ and 
all accessible excited states $|a'\rangle$. 
This Rabi-oscillation of the K-shell electron between the bound states of the Ne$^{+2}$ ion will induce an ac-Stark splitting  
manifested in the kinetic energy of the ejected Auger-electron. We should note that no post-collision 
interaction with the photo-electron is taken into account, as the energy of the photoelectron is too large for such an effect to contribute 
in the Auger's electron spectrum (and vice-versa). To complete the picture, the excited states $| a' \rangle$ decay either through
 an additional Auger process ($\Gamma_{a'}$) or through further ionization by absorbing one more photon ($\gamma_{a'}$). 
Finally, direct creation of a double core-hole Ne$^{+2}$ of the neutral, through photoabsorption, is not considered since the double K-shell 
ionization energy is 1863 eV \cite{pelicon:2000}. 


We discuss below the manner in which the AES is modified due to the Rabi-coupling of the Ne$^{+2}(1s^{-1}-3p)$ states and 
study it's behaviour in a quantitative manner.
To facilitate the interpretation of the results, we note at this point that for a modification of the Auger line to appear, 
many Rabi-oscillations should occur within the relevant Auger lifetime. 
Equivalently, in the energy domain, it is required that the energy separation of the Auger-line splitting 
(roughly equal to Rabi coupling strength) should be larger than the Auger decay width or the x-ray bandwidth, whichever is larger. 
Of course the detailed properties of the system and the field will matter as well,  however the rule of thumb, as expressed above will be in general true.

The structure of the paper is as follows. In section II we present the theoretical formulation and develop the decription of the processes 
in terms of a time-dependent density matrix 
system of equations. In section III we show some of the results of our present study in the case of a single-mode coherent field. In the final 
section (Sec. IV) we summarize our findings and discuss, very briefly, two aspects of the problem that are necessary to put the description of the problem 
closer to the actual experimental conditions, namely the  fluctuations present in a FEL field and the field's spatial dependance. 


%


\begin{center}
\begin{table}[t]
\begin{tabular}{|c|c|c|c|}
\hline
$|a' \rangle$ & $E_{a'}$ (eV) & Ne$^{+2}(1s^12s^2)$ & $ gf_{aa'}(\times 10^{-2}$) \\
\hline
\hline
1 &  907.75  & $(2p^4, ^1\!D)^2\!D(3p^1)^1\!P_1$ & 2.3338\\
\hline
2 &  907.90  & $(2p^4, ^3\!P)^2\!P(3p^1)^3\!P_1$ & 0.20991\\
\hline
3 &  908.06  & $(2p^4, ^1\!D)^2\!D(3p^1)^1\!F_3$ & 8.1881\\
\hline
4 &  908.48  & $(2p^4, ^3\!P)^2\!P(3p^1)^3\!D_3$ & 0.13141\\
\hline
5 &  908.51  & $(2p^4, ^1\!D)^2\!D(3p^1)^3\!D_2$ & 0.23322\\
\hline
6 &  908.49  & $(2p^4, ^3\!P)^2\!P(3p^1)^1\!D_2$ & 4.4888\\
\hline
7 &  908.78  & $(2p^4, ^1D)^2D(3p^1)^1D_2$ & 1.2714\\
\hline
\end{tabular} 
\caption{The above table lists the transitions from the $| a \rangle $ = Ne$^{+2}$($^1$D$_2)$ ground state to 
its excited states Ne$^{+2}(1s^{-1}-3p)$ around the photon frequency $\omega =$ 908 eV. The fourth column shows 
the corresponding oscillator strengths. The data are calculated using the Cowan suites of codes \cite{cowan:code}.}
\end{table}
\end{center}
\section{Theoretical formulation of the density matrix equations}
The density operator of the system is obtained in the basis of $|G \rangle, |I \rangle, |A \rangle, |A^\prime \rangle, |R \rangle, |F_i \rangle, i = 1,2$ states.  
The state $| G \rangle $, with energy $E^{(g)}$, represents the neon ground state. The state
$| I \rangle = | i ; \varepsilon_i \rangle$, with energy $E_i = E^{(i)} + \varepsilon_i$, represents the 
K-shell hole Ne$^{+}$ (state $|i \rangle$ with energy $E^{(i)}$) and the photo-ejected electron $| \varepsilon_i \rangle$ having 
kinetic energy $\varepsilon_i$. 
The state $|A \rangle = |a;\varepsilon_a,\varepsilon_{ia} \rangle $, with energy $E_a = E^{(a)} +\varepsilon_a + \varepsilon_{ia}$, 
represents the Ne$^{+2}$ ion in its ground state (state $|a \rangle$ with energy $E^{(a)})$ with an ejected Auger-electron of kinetic energy 
$\varepsilon_a$ and the photoelectron having now kinetic energy $\varepsilon_{ia}$. 
Similarly, the state $ |A' \rangle = |a';\varepsilon_{a'}, \varepsilon_{ia'} \rangle$, 
with energy $E_{a'} = E^{(a')} +\varepsilon_{a'} + \varepsilon_{ia'} $, represents the excited state of Ne$^{+2}$ (state $|a^\prime \rangle $ 
with energy $E^{(a')}$) with the Auger electron having kinetic energy $\varepsilon_{a'}$ and the photoelectron having 
now kinetic energy $\varepsilon_{ia'}$. 
 It should be noted that in the definition of the photoelectron and Auger-electron states 
the appropriate angular momentum quantum numbers, as they result from electric dipole and Auger transition rules,  are included.
In addition, we also take into account the possibility of the involvement of further decay modes. In the present case, the K-shell hole 
Ne$^{+}$ $|i \rangle$ can decay through fluorescence to Ne$^+ (1s^22s^22p^5)$, denoted here as $|R \rangle$. 
Moreover, the excited states $|A'\rangle$ may further decay either through an Auger transition to  Ne$^{+3}(1s^22s^22p^3)$ 
denoted as $| F_1 \rangle$ or through further photoionization to hollow K-shell Ne$^{+3} (1s^12s^22p^4)$, denoted as $|F_2 \rangle$. 
The equations of motion for the density matrix elements are obtained from the Liouville equation $i\dot{\rho}(t) = [H(t),\rho(t)]$ 
with $\hat{H}(t) = \hat{H}^{0} + \hat{V} + \hat{D}(t)$, $\hat{H}^0$ being the field-free Hamiltonian of neon, $\hat{V}$ 
the electron-electron interaction operator and $\hat{D}(t)$ the x-ray field-atomic dipole interaction operator. 
Inserting the above states into the Liouville equation we obtain:
\begin{eqnarray*}
\dot{\rho}_{GG}(t)                                   & = &  2 Im\sum_{I} \nspace \int D_{GI} \rho_{IG} ,      \\
\dot{\rho}_{II}(t)                     & = & 2 Im \left[D_{IG} \rho_{GI} \right]
                                                            + 2 Im \sum_{A}   \nspace \int  V_{IA} \rho_{AI}  \\
                                                           & +& 2 Im \sum_{R} \nspace \int D_{IR} \rho_{RI}   \\
\dot{\rho}_{AA}(t)    & = &    2  Im \left[ V_{AI} \rho_{IA} \right]
                                                            + 2 Im \left[ D_{AA'} \rho_{A'A} \right]                    \\
\dot{\rho}_{A'A'}(t)    & = &  - 2 Im \left[ D_{AA'} \rho_{A'A} \right] + 2 Im \sum_{F_1} \nspace \int V_{A'F_1} \rho_{F_1A'} \\
 &+& 2 Im \sum_{F_2} \nspace \int D_{A'F_2} \rho_{F_2A'}\\
i\dot{\rho}_{AA'}(t)   &=& E_{AA'} \rho_{AA'} + D_{AA'}(\rho_{A'A'} - \rho_{AA}) + V_{AI} \rho_{IA^\prime}\\
                                                   & -&  \sum_{F_1} \nspace \int \rho_{AF_1} V_{F_1A'} 
                                                    -    \sum_{F_2} \nspace \int \rho_{AF_2} D_{F_2A'} 
\\
i\dot{\rho}_{GI}(t) & = & ... \nonumber \\
...&...&...
\end{eqnarray*}
In the above expressions $\rho_{KL}, K,L = G,I,A,A^\prime, R,F_1,F_2$ are the density matrix elements of the involved states 
while $D_{KL}$ and $V_{KL}$ represent electric dipole and Auger (intra-atomic) transitions between the states $K,L$, respectively. 
More specifically, the quantities $D_{GI},V_{IA},D_{IR}, D_{AA^\prime}, V_{A^\prime F_1}, D_{A^\prime F_2}$ represent multielectron electric dipole 
($D$) and Auger ($V$) transition matrix elements. Within the present context we do not take into account any post-collision effects between 
the photo- and Auger-electrons as their contribution are expected to be negligible for the considered kinetic energies. 
This assumption allows for a simplification of the transition matrix elements as for example for 
$V_{IA} = \langle i, \varepsilon_i|\hat{V}| a, \varepsilon_{a}, \varepsilon_{ia} \rangle$ 
which reduces to $V_{IA} = \langle i | \hat{V} | a \rangle \langle \varepsilon_{ia} | \varepsilon_i \rangle = 
V_{ia} \delta(\varepsilon_{ia} - \varepsilon_{i})$. Along the same lines the dipole transition $D_{AA^\prime} = \langle a, \varepsilon_{a}, \varepsilon_{ia} | \hat{D} | a^\prime, \varepsilon_a', \varepsilon_{ia'} \rangle$, is approximated as  $D_{AA'} = \langle a |\hat{D}| a'\rangle  \langle \varepsilon_{a} | \varepsilon_{a'} \rangle \langle \varepsilon_{ia} | \varepsilon_{ia'} \rangle = d_{aa'} 
\delta(\varepsilon_a-\varepsilon_{a'})\delta(\varepsilon_{ia}-\varepsilon_{ia'})$. A detailed discussion of the dimensional reduction of 
these special kind of continuum-continuum matrix elements can be found in the appendix of Ref. \cite{hanson:1997}.
The summations involved here imply integration over the appropriate continua.  
As the total number of independent equations is 28, we do not present the explicit expressions for the evolution 
of the remaining density matrix elements as they are not essential at this stage. 

The density matrix equations, are a system of coupled integro-differential equations which are 
not amenable to an easy solution even by numerical means as it includes integration over multidimensional continua. 
It is thus our purpose here to transform the above system of equations into a more tractable form. To this end, we adiabatically eliminate 
the density matrix elements which are involved in the integrations over the respective continua of the states. 
The procedure for adiabatically eliminating these continua is a standard technique applied to describe the influence of a system 
with infinite degrees of freedom on to a system with a small number of degrees and appears in many 
different contexts (see e.g. \cite{louisell:1990}). 
Here, the reduced system is the one described by $|G \rangle, |I \rangle, | A \rangle$  and $|A^\prime \rangle$ while $|R \rangle, |F_i \rangle, i =1,2$ 
represent the dissipative environment. Within the present context of atomic continua, some of the details can also be found in 
\cite{lambropoulos:1983}. To proceed further, the radiation field is expressed as 
${\bf E}(t) = \hat{e}(\mathcal{E}(t)e^{i\omega t} + \mathcal{E}^{\star}(t)e^{-i\omega t})/2$ with $\hat{e}$ its polarization vector and 
we transform to slowly varying variables by defining 
$\sigma_{kl}= \rho_{KL}e^{-i n\omega t}, n=0,\pm 1,\pm 2, \pm 3$, where $n$ is chosen so that $n \omega$ has the closest possible value to 
$E_K-E_L$.
With the latter transformation we remove from the coherences the fast oscillation part of their evolution due to the frequency 
of the field (this is justified since for a frequency of 1 keV the field period is of the order of 4 as and all other time scales set 
by photoionization and Auger widths are of the order of  1 fs $ \sim 1000 $ as). 
Given that the radiation is in the form of a pulse we have kept the slowly varying envelope $\mathcal{E}(t)$ which in addition may describe 
the stochastic properties of the field under consideration. This is however a problem which requires special care and postpone its discussion for now.  
In the present case, we assume a fully coherent single-mode, Fourier transform-limited, field. 
Then, by employing the rotating wave approximation (RWA) and keeping only the terms proportional to the first-order of the electric field, 
after tedious but straightforward manipulation we end up to the following set of equations for the reduced density matrix elements 
\begin{widetext}
\begin{subequations}
 \label{eq:sigma_t}
\begin{eqnarray}
\dot{\sigma}_{gg}(t)                                 & = & - \gamma_g \sigma_{gg},      
\label{eq:rho_1}                                                              \\
\dot{\sigma}_{ii}(\varepsilon_i,t)                   & = & - \Gamma_i\sigma_{ii}  + Im\left[ \Omega^{\star}_{ig} \sigma_{gi}\right], 
 \label{eq:rho_2}            \\
\dot{\sigma}_{aa}(\varepsilon_i, \varepsilon_a,t)    & = & - Im\left[\Omega_{a'a}^{\star}\sigma_{aa'}\right] + 2Im\left[V_{ai}\sigma_{ia}\right],  \label{eq:rho_3} \\
\dot{\sigma}_{a'a'}( \varepsilon_i, \varepsilon_a,t) & = & - \bar{\gamma}_{a'}\sigma_{a'a'}  + Im\left[\Omega_{a'a}^{\star}\sigma_{aa'}\right],   \label{eq:rho_4} \\
i\dot{\sigma}_{aa'}(\varepsilon_i, \varepsilon_a,t)  & = &   (E_{aa'}+\omega-i\frac{\bar{\gamma}_{a'}}{2})\sigma_{aa'}
                                                           + \frac{\Omega_{aa'}}{2} (\sigma_{a'a'}- \sigma_{aa})  + V_{ai}\sigma_{ia'},         
\label{eq:rho_5}                                                         \\
i\dot{\sigma}_{gi}(\varepsilon_i,t)                  & = &   (E_{gi} + \omega-i\frac{\gamma_{g} + \Gamma_i }{2})\sigma_{gi} 
                                                           - \frac{1}{2}\Omega_{gi}\sigma_{gg},                                            
\label{eq:rho_6}                                                               \\
i\dot{\sigma}_{ia}(\varepsilon_i, \varepsilon_a,t)   & = &   (E_{ia}-i\frac{\Gamma_i}{2})\sigma_{ia} + \frac{1}{2}\Omega^{\ast}_{ig}\sigma_{ga} 
                                                           - \frac{1}{2}\Omega^{\ast}_{a'a}\sigma_{ia'} - V_{ia}\sigma_{ii},       
\label{eq:rho_7}                                                                 \\
i\dot{\sigma}_{ia'}(\varepsilon_i, \varepsilon_a,t)  & = &   (E_{ia'}+\omega-i\frac{\Gamma_i+\bar{\gamma}_{a'}}{2})\sigma_{ia'}
                                                           + \frac{1}{2}\Omega^{\ast}_{ig}\sigma_{ga'} -\frac{1}{2}\Omega_{aa'}\sigma_{ia}, 
\label{eq:rho_8}                                                      \\
i\dot{\sigma}_{ga}(\varepsilon_i, \varepsilon_a,t)  & = &    (E_{ga}+\omega-i\frac{\gamma_{g}}{2})\sigma_{ga} 
                                                           - \frac{1}{2}\Omega^{\ast}_{a'a}\sigma_{ga'} - V_{ia}\sigma_{gi},                     
\label{eq:rho_9}                                                  \\
i\dot{\sigma}_{ga'}(\varepsilon_i, \varepsilon_a,t)  & = &   (E_{ga'}+2\omega - i\frac{\gamma_{g} +\bar{\gamma}_{a'} }{2})\sigma_{ga'} 
                                                           - \frac{1}{2}\Omega_{aa'}\sigma_{ga},
                                                           \label{eq:rho_10}
\end{eqnarray}
\end{subequations}
\end{widetext}
where $\Omega_{gi}(\varepsilon_i,t) =  \langle g | \hat{D} | i, \varepsilon_i \rangle \mathcal{E}(t)$ and 
$\Omega_{aa'} (t)= d_{aa^\prime} \mathcal{E}(t)$. 
With $\gamma_g(t) = 2 \pi \int d \varepsilon_i |\Omega_{gi}(\varepsilon_i,t)|^2$ we denote the photoionization width of the neon ground states relative to the Ne$^+$ K-shell hole state $|i \rangle$, while 
$\bar{\gamma}_{a'} = \Gamma_{a'} + \gamma_{a'}(t)$ is the sum of the Auger decay width to states $| F_1 \rangle$ ($\Gamma_{a'}$) and 
the photoionization width to states $|F_2 \rangle $ ($\gamma_{a'}$) of the excited states $| a' \rangle $. In addition, 
$\Gamma_i = \Gamma_{ia} + \gamma_r$, is the sum of the decay width of the ionic state $| i \rangle$ through Auger decay 
to states $|a \rangle $ and through fluorescence to states $|R \rangle$. 
The quantity $V_{ia}$ represents the strength of the Auger transition of the hole state $| i \rangle$ to the particular ionic ground state $|a\rangle$. Therefore, 
the quantity $\Gamma_{ia}$  is expressed as $\Gamma_{ia}= 2 \pi |V_{ia}|^2$. 
Finally the energy differences in the above equations now include all the shifts associated with the Auger and dipole couplings of the relevant states 
with continuum states $E_{kl} \equiv E_k + S_k - (E_l + S_l), k,l = g, i, a^\prime $ with $S_i = S_{ia} + S_{ir}$ and 
$S_{a^\prime} = S_{a^\prime f_1} + S_{a^\prime f_2}$. The exact definition of the shifts and widths are as below:
\begin{eqnarray*}
S_{g} - i\frac{\gamma_g}{2}             &=& 
\lim_{\eta \rightarrow 0 }\int \!\!dE_I \frac{|D_{GI}|^2}{E_G + \omega - E_{I} + i\eta}, \\
S_{ia} - i\frac{\Gamma_{ia}}{2}          &=& 
\lim_{\eta \rightarrow 0 }\int \!\!d E_A \frac{|V_{IA}|^2}{E_I - E_A + i\eta}, \\
S_{ir} - i\frac{\Gamma_{ir}}{2}          &=& 
\lim_{\eta \rightarrow 0 }\int \!\!dE_R \frac{|D_{IR}|^2}{E_I - E_R + i\eta}, \\
S_{a^\prime f_1} - i\frac{\Gamma_{a^\prime}}{2} &=& 
\lim_{\eta \rightarrow 0 }\int \!\!dE_{F_1} \frac{|V_{A'F_1}|^2}{E_{A'} - E_{F_1} + i\eta},\\
S_{a^\prime f_2} - i\frac{\gamma_{a^\prime}}{2} &=& 
\lim_{\eta \rightarrow 0 }\int dE_{F_2} \frac{|D_{A'F_2}|^2}{E_{A'} + \omega- E_{F_2} + i\eta},
\end{eqnarray*}
where use of the well-known formula $\lim_{\eta \rightarrow 0} 1/(x+i\eta) = \mathcal{P}(1/x)-i\pi \delta(x)$ must be made to split up 
the integrals into their real and imaginary parts.  

At this stage, a working set of equations are established and their numerical solution is feasible, provided that all 
the dynamical parameters of the problem have been calculated before-hand. The approximations leading to this set of equations 
require careful examination of the appropriate range of radiation intensities. One approximation is to assume that the relevant continua are 
smooth around the energies of the dressed bound states within an energy range comparable with the Rabi-coupling matrix element. This 
requires that transitions close to ionization thresholds should not be considered. In addition, in the derivation procedure, 
we have ignored terms proportional to the second order of the field, such that $\Omega_{aa^\prime} < 1$. The latter approximation, 
given the matrix element $d_{aa^\prime}  \sim 0.06$ a.u., will restrict the range of the intensities where the working equations 
are applicable below to $4 \times 10^{18}$ W/cm$^2$.

The system of equations that have been derived must, simultaneously, be numerically integrated for all different photo-electron and 
Auger-electron kinetic energies, so as to provide the populations for $\sigma_{ii}(\varepsilon_i), \sigma_{aa}(\varepsilon_a,\varepsilon_i),
\sigma_{a'a'}(\varepsilon_a,\varepsilon_i)$ at infinite times. Since in our case we are only interested in the Auger-kinetic energy spectrum and ionization yields 
regardless of the state of the photoelectron, we must integrate the final populations over the photo-electron kinetic energies and determine 
the following probabilities $\sigma_{ii} = \int d\varepsilon_i  \sigma_{aa}(\varepsilon_i), 
\sigma_{aa}(\varepsilon_a) = \int d\varepsilon_i  \sigma_{aa}(\varepsilon_a,\varepsilon_i),
\sigma_{a'a'}(\varepsilon_a) = \int d\varepsilon_i  \sigma_{a'a'}(\varepsilon_a,\varepsilon_i)$. 
An alternative and more economical way of obtaining the same results is to derive a coarse-grained version of the present equations for 
these reduced, averaged over the photoelectron energies, density matrix elements. In addition, the reduced set of the density 
matrix equations is amenable to further manipulation as it allows for the derivation of analytical expressions for long pulses or 
their averaging for stochastic pulses. 
Thus, the new reduced set of equations, is obtained by first setting all the derivatives 
of the coherences equal to zero [except $\dot{\sigma}_{aa'}(\varepsilon_a, \varepsilon_i,t)$] and then integrating over 
the photoelectron energy $\varepsilon_i$.  
To demonstrate the reasoning of setting the derivatives of the coherences to zero, we work out the evolution equation of $\sigma_{gi}(\epsilon_i,t)$ coherence 
[Eq. (\ref{eq:rho_6})]. We integrate Eq. (\ref{eq:rho_6}) in an interval $t,t+\tau$ with $\tau << \Omega^{-1}_{aa^\prime}, \gamma^{-1}_g, \Gamma^{-1}_i$ 
and we obtain:
\begin{widetext}
\begin{displaymath} 
i[ \sigma_{gi}(\varepsilon_i,t+\tau) - \sigma_{gi}(\varepsilon_i,t)] =  (E_{gi}(\tau) + \omega-i\frac{\gamma_{g}(\tau) + \Gamma_i }{2})
\int_t^{t+\tau} \!\!dt^\prime \sigma_{gi}  (\varepsilon_i,t^\prime)  - \frac{\Omega_{gi}(\varepsilon_i, \tau)}{2} \int_t^{t+\tau}\!\!dt'\sigma_{gg}(t^\prime).
\end{displaymath}
\end{widetext}
The ionization width $\gamma_g(\tau)$, the dipole $\Omega_{gi}(\varepsilon_i,\tau)$ and the $S_g(\tau), S_i(\tau)$ ac-Stark shifts that are included in the definition of $E_{gi}(\tau)$ were removed from the integral as their value doesn't change much 
between $t$ and $t+\tau$ as a result of the slowly varying trasformation of the variables. Since it will always be 
$|\sigma_{gi}(\varepsilon_i,t)| << \sigma_{gg}(t)$, we can neglect the left hand side and obtain the 'coarse grained' time average 
of $\sigma_{gi}(\varepsilon_i,t)$ as: 
\begin{equation}
\bar{\sigma}_{gi}(\varepsilon_i,\tau) = \frac{\Omega_{gi}(\varepsilon_i,\tau)/2 }{ E_{gi}(\tau) + \omega-i(\gamma_{g}(\tau) + \Gamma_i)/2 }\bar{\sigma}_{gg}(\tau), 
\end{equation}
where $\bar{\sigma}_{gi}(\varepsilon_i,\tau) \equiv \int_t^{t+\tau} \!\!dt^\prime \sigma_{gi}  (\varepsilon_i,t^\prime)/\tau$ and 
$ \bar{\sigma}_{gg}(\tau) \equiv\int_t^{t+\tau}\!\!dt'\sigma_{gg}(t^\prime)/\tau$.  Thus, by setting the derivative of the coherence 
to zero, effectively, leads to a coarse-grained value for the coherence which follows adiabatically the ground state population. 
At this stage, integrating Eq. (\ref{eq:rho_2}) over time (in an interval  $[t,t+\tau]$) and the photoelectron energy $\varepsilon_i$ we obtain:
\begin{widetext}
\begin{eqnarray}
\dot\sigma_{ii}(\tau) &=& - \Gamma_i \sigma_{ii}(\tau) 
+ Im\!\!\int\!\!\!  d\varepsilon_i \frac{ |\Omega_{gi}(\varepsilon_i,\tau) |^2/2}{ E_{gi}(\tau) + \omega-i(\gamma_{g}(\tau) + \Gamma_i)/2}  \sigma_{gg}(\tau) \nonumber \\
&=&  - \Gamma_i \sigma_{ii}(\tau)  + \gamma_g(\tau) \sigma_{gg}(\tau), 
\label{eq:sigma_ii}
\end{eqnarray}
\end{widetext}
where $\sigma_{ii}(\tau) = \int d\varepsilon_i \bar{\sigma}_{ii}(\varepsilon_i,\tau)$ and $\bar{\sigma}_{ii(\varepsilon_i,\tau)} = \int_t^{t+\tau} dt^{\prime} \sigma_{ii}(\varepsilon_i,t^\prime) $. To evaluate the integral we have assumed 
that $\Omega_{gi}(\varepsilon_i,\tau)$ is smooth over an energy range equal to the radiation's bandwidth (far from resonance structures in the continuum or ionization thresholds). Then by expressing $E_{gi}$ as 
$E_{gi}(\tau) = E^{(g)} +S_g(\tau) - E^{(i)} -S_i(\tau) -\varepsilon_i$ we have  $Im\int d\varepsilon_i|\Omega_{gi}(\varepsilon_i,\tau)|^2/(E_{gi}(\tau) + \omega-i(\gamma_{g}(\tau) + \Gamma_i)/2) \sim \pi |\Omega_{gi}(\varepsilon_i = E^{(i)}+S_i-  E^{(g)} -S_g , \tau )|^2=2\gamma_g(\tau)$.

As the derivation is quite long and detailed for the remaining coherences, we give here only the final result for the reduced (coarse-grained) 
set of density matrix equations. These reduced equations are obtained working along similar lines as for the derivation of  Eq. (\ref{eq:sigma_ii}). 
In this derivation, we ingore the photoionization of the excited $| a^\prime \rangle$, $ (\gamma_{a'}=0)$ as for the photon energies and the intensities 
considered is expected to be much less than the Auger decay transition, represented by $\Gamma_{a'}$. 
Therefore, after setting all the derivatives of the coherences equal to zero [Eqns. (\ref{eq:rho_6}), (\ref{eq:rho_7}), (\ref{eq:rho_8}), 
(\ref{eq:rho_9}), (\ref{eq:rho_10})], we solve for the coherences and substitute their values into Eqns (\ref{eq:rho_3}), (\ref{eq:rho_4})  and (\ref{eq:rho_5}). 
Then we integrate our equations over the photoelectron's kinetic energy and obtain the coarse-grained (also changing $\tau \rightarrow t$) set of density matrix equations, 
\begin{widetext}
\begin{subequations}
\label{eq:sigma_t_coarse_grained}
\begin{eqnarray}
\dot{\sigma}_{gg}(t)& =& - \gamma_g \sigma_{gg}, \label{eq:sigma_1}\\
\dot{\sigma}_{ii}(t)& =  & -\Gamma_i\sigma_{ii}  + \gamma_g \sigma_{gg}, \label{eq:sigma_2}  \\
\dot{\sigma}_{aa}(\varepsilon_a,t)& = & - Im\left[\Omega_{a'}^{\star}\sigma_{aa'}\right]
 + Im\left[( \Delta_{a}+\delta_{a'} - i\frac{\Gamma_i+\Gamma_{a'}}{2}  )
(\Omega_{a'}^{+}- \Omega_{a'}^{-})\right]\sigma_{ii}, \label{eq:sigma_3} \\
\dot{\sigma}_{a'a'}(\varepsilon_a,t)& = & -\Gamma_{a'}\sigma_{a'a'}  + Im\left[\Omega_{a'}^{\star}\sigma_{aa'}\right],   \label{eq:sigma_4}\\
i\dot{\sigma}_{aa'}(\varepsilon_a,t) &=&(\delta_{a'}-i\frac{\Gamma_{a'}}{2})\sigma_{aa'} 
- \frac{\Omega_{a'}}{2} (\sigma_{a'a'}- \sigma_{aa}) 
+ 
\frac{\Omega_{a'}}{4}( \Omega_{a'}^{+} - \Omega_{a'}^{-}) \sigma_{ii}. \label{eq:sigma_5}
\end{eqnarray}
\end{subequations}
\end{widetext}
The dynamics of the process are governed by the ionization width of the neutral target $\gamma_g(t)$, the core Rabi-coupling $\Omega_{a'}(t)$ 
the intra-atomic decay rates $\Gamma_i,\Gamma_{a'}$, the Auger-field induced couplings $\Omega_{a'}^{\pm}(t)$ and the Auger $\Delta_a(t) = E_i - (E_a + \varepsilon_a)$ and  field $\delta_{a'}(t)=(E_{a}+\omega) - E_{a'}$ detunings.
 Note that for notational simplicity we denote the core Rabi coupling as 
$\Omega_{a'} = \Omega_{aa^\prime} = \Omega_{a^\prime a}$. 
We have defined $\Omega_{a'}^{\pm}(t) = 2|V_{ia}|^2/(\Delta_{a'}^{\pm}\bar{\Omega}_{a'}) $ with 
$\Delta_{a'}^{\pm}(t) = \epsilon_a -  \epsilon_a^{\pm} + i \gamma_a^{\pm}(t)/2 $,  
$\epsilon_a^{\pm}(t) = \epsilon_a^{(0)} + [\delta_{a'} \mp \bar{\Omega}^{(r)}_{a'}(t) ]/2$ and 
$\gamma_a^{\pm}(t) =  \Gamma_i + [\Gamma_{a'} \pm \bar{\Omega}^{(i)}_{a'}(t)]/2$. The real quantities $\bar{\Omega}^{(r)}_{a'}$ and 
$\bar{\Omega}_{a'}^{(i)}$ are defined in terms of the generalized Rabi frequency:
\begin{equation}
\bar{\Omega}_{a'}(t) = \bar{\Omega}^{(r)}_{a'} + i \frac{\bar{\Omega}_{a'}^{(i)}}{2} 
= \sqrt{(\delta_{a'} - i\frac{\Gamma_{a'} }{2})^2 + 4 |\Omega_{a'}|^2}.
\label{eq:rabi}
\end{equation}

The AES, at detection time, is obtained by adding the contributions from the ground $|a \rangle$ and excited  $|a'\rangle$ states 
of the doubly-ionized neon: $ S(\varepsilon_a) =\sum_{j=a,a'} \int_{-\infty}^{+\infty} dt^{\prime} \dot{\sigma}_{jj}(\varepsilon_a,t^{\prime}) $. 
The populations of the states $|a\rangle$ and $| a^\prime\rangle$ are obtained as $p_{jj}(t) = \int^{+\infty}_{-\infty} dt \int d\varepsilon_i \dot{\sigma}_{jj}(\varepsilon_a,t^{\prime}),\quad j = a,a^\prime $.
Although we numerically integrate the above system of density matrix equations Eqns (\ref{eq:sigma_t_coarse_grained}) the Auger spectra can 
also be cast in analytical form to a very good approximation. For example, for a non-decaying excited states $(\Gamma_{a'}=0)$ the generalized Rabi 
frequency becomes a purely real quantity $\bar{\Omega}_{a^{\prime}} = \sqrt{\delta^2_{a^\prime} + 4 |\Omega_{a^\prime}|^2}$ while the AES become independent on the coherence
evolution $\sigma_{aa^\prime}$. The analytical approximation consists of considering a pulse of constant amplitude $\mathcal{E}(t)$ which 
turns the coarse-grained system to a system of ordinary differential equations with constant coefficients. Then the Rabi frequency, 
ac-Stark level shifts and the ionization widths become independent on time and in the expression for the AES only the population 
of $|i \rangle$ is time-dependent:
\begin{eqnarray}
 S(\varepsilon_a) &=& \int_{-\infty}^{+\infty} dt\left[ \dot{\sigma}_{aa}(\varepsilon_a,t) + \dot{\sigma}_{a'a'}(\varepsilon_a,t)\right] \nonumber \\
                  &=& Im[( \Delta_{a}+\delta_{a'} - i\frac{\Gamma_i}{2}) (\Omega^{+}_{a'}-\Omega^{-}_{a'} )] \int_{-\infty}^{+\infty} dt\sigma_{ii}(t). \nonumber 
\end{eqnarray}
Solving the Eqns (\ref{eq:sigma_1}) and (\ref{eq:sigma_2}) for $\sigma_{ii}(t)$ we find that 
$\int_{-\infty}^{+\infty} \! dt \sigma_{ii}(t) = 1/\Gamma_i$ and after some algebra we end up to the following analytical expression for the AES:
\begin{widetext}
\begin{equation}
S(\varepsilon_a)  = \frac{\Gamma_{ia}}{4\pi}
\left[
\frac{1-\delta_{a'}/\bar{\Omega}_{a'}}{(\varepsilon_a-\varepsilon^{(0)}_a  - \frac{\delta_{a'}-\bar{\Omega}_{a'}}{2})^2 + \frac{\Gamma_i^2}{4}} + 
\frac{1+\delta_{a'}/\bar{\Omega}_{a'}}{ (\varepsilon_a-\varepsilon^{(0)}_a - \frac{\delta_{a'}+\bar{\Omega}_{a'}}{2})^2+ \frac{\Gamma_i^2}{4}}
\right]. 
\label{eq:aes}
\end{equation}
\end{widetext}
We should note here, that the predictions of the analytical expression differ from the numerical solution in that it doesn't 
include the transient effects of the physical process which are expected to occur at times of the order of $1/\Gamma_i$:

\section{Results}

In Fig. 2 we show the effect of the field strength in the Auger spectra for  $\omega=908.06$ eV and 
assume a pulse envelope $\mathcal{E}(t) = \sin^2(\pi t /T)$, with $T$ being the total pulse duration equal to about 20 times the Auger life time 
$\sim 48.8 $ fs (FWHM = 24.4 fs). More specifically in all the following calculations the field included 10754 cycles.
The ionization width of the neutral is given by $\gamma_g(t) = 4.375\times10^{-4} \mathcal{E}^2(t)$ a.u..
In the calculation we only include the  $| a^\prime \rangle $ =  $|$Ne$^{+2}(1s^{-1}-3p),$ $^1$F $\rangle $ 
excited state [state 3 of table 1] and assume its decay width to be zero ($\Gamma_{a'}=0$).
For this, on-resonance process ($\delta_{a'}=0$),  the Rabi interaction energy is found to be $\Omega_{a^\prime} = 0.061 \mathcal{E}(t)$ a.u..
According to Eq. (\ref{eq:aes}), for peak intensities of $I_0 = 1.0\times 10^{15}, 1.0\times 10^{16}, 3.51\times 10^{16}, 1.0\times 10^{17} $ 
and $3.51\times 10^{17}$ W/cm$^2$ peak separations (of equal height) of about $0.28, 0.88, 1.66, 2.79$  and $5.25$ eV, respectively, 
should be expected for the AES. Particularly, for the lowest intensity of $I_0 = 1.0\times 10^{15} $ the peak separation is comparable 
to the Auger decay width (0.27 eV), thus the separation is hardly seen. The latter values coincide with the full numerical solutions, shown in Fig 2. 
In general, according to the above formula, for on-resonance conditions and long pulses (compared to $1/\Gamma_i$) 
a change of the splitting by a factor around 3.2 should be expected, for a change of the peak intensity by one order of magnitude. 
For off-resonance conditions we have two unequal peaks with energy separation again determined from the generalized 
Rabi-frequency $\bar{\Omega}_{a^\prime}$ and relative height determined from the field detuning $\Delta_{a'}$.


\begin{figure}[h!]
\includegraphics[scale=0.3]{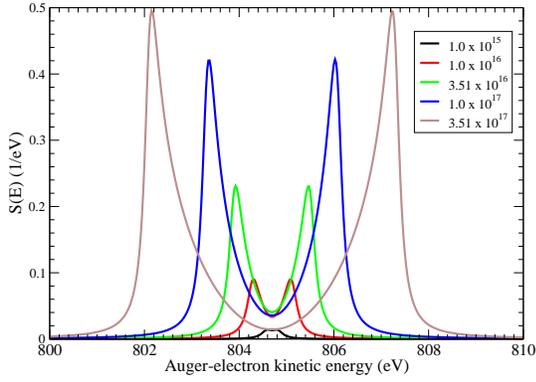}
\caption{(Color online) Variation in the AES as the peak intensity of the pulse increases. 
The field has total duration about 48.8 fs (FWHM = 24.4 fs) and the photon frequency chosen equal to 908.06 eV, 
while $\Gamma_{a'} = 0$. The inset label give the peak intensity values in units of W/cm$^2$.}
\label{fig:fig2}
\end{figure}
Next we turn to the case where the decay channels of the excited core states $|a^\prime \rangle$ are present. 
We have found that for the intensities and photon energy considered, further photoionization of $|a^\prime \rangle $ is a much 
weaker channel compared with a RAS transition to Ne$^{+3}$ \cite{cowan:code, bhalla:1973}. 
We take the RAS width to be a large fraction of Ne$^{+3}(1s2p^22p^4) \rightarrow$ Ne$^{+3}(1s^22p^22p^3)$ decay width 
and assume $\Gamma_{a'} = 0.156 $ eV \cite{bhalla:1973}. 
In this case the life time of these excited ionic states $|a^\prime \rangle$ is about 4.12 fs, comparable to the Auger decay 
life time of interest here ($\sim$ 2.45 fs) but still much shorter than the pulse duration. 
Given that the Rabi coupling saturates the Ne$^{+2}(1s^{-1}-3p)$ transition very quickly (in the sense that their populations 
are almost equalized) it can be expected that a large portion of the population will very quickly decay to the Ne$^{+3}$ ion. 
This behaviour is shown in Fig. 3 where the final populations in Ne$^{+2}$ and Ne$^{+3}$ are plotted as a function of the peak intensity 
of the applied x-ray field. Thus for on-resonance cases (solid lines) we see that the majority of the population goes 
into Ne$^{+3}$ ion for all intensities considered. On the other hand, if we choose to detune the FEL to a photon energy of 904.06 eV 
(dashed lines), efficient population of the excited ionic state $|a' \rangle$ is prohibited.
This also causes an effective increase in the magnitude of the generalized Rabi frequency according to Eq. (\ref{eq:rabi}). 
In that case the situation changes dramatically. The relative population ratio is reversed for low intensity fields ($ < 10^{17}$ W/cm$^2$), 
with the Ne$^{+2}$ dominating up to higher intensities where the ratio starts to decline in favor of the Ne$^{+3}$ yield. 
Thus, it appears that a careful combination of the intensity and the photon energy of the field can control the relative 
populations of the triply and doubly ionic species of neon. 
\begin{figure}[t]
\includegraphics[scale=0.3]{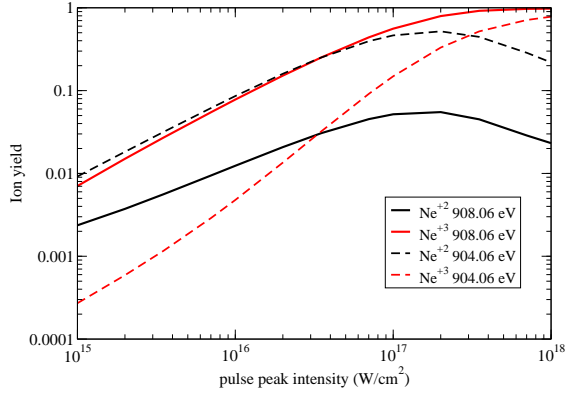}
\caption{(Color online). Ionization yields versus the peak intensity of the x-ray field. Field parameters as in Fig. 2 
and $\Gamma_{a'} = 0.156 $ eV. Solid curves refer to $\omega = 908.06$ eV while the dashed ones to $\omega = $ 904.06 eV.}
\label{fig:fig3}
\end{figure}
It is worth to note here that for the present long pulses and for relatively strong fields ($ > 3.5 \times 10^{16}$ W/cm$^2$), 
the depletion of the neutral neon might be significant. For example, for $I_0 = 3.5 \times 10^{16}$ W/cm$^2$ the remaining 
neutral neon is 0.718 while for ten times stronger field $I_0 = 3.5 \times 10^{17}$ W/cm$^2$ decreases to 0.036. 
Furthermore, it should also be noted that ionic species higher than triply and doubly ionized neon are not expected to contribute 
significantly as two- or multi- photon absorption is an unlikely ionization channel, given their ionization potentials and the large 
photon energy. Finally, a small fraction of Ne$^{+1}$ is produced through the fluorescence of the metastable K-shell hole Ne$^{+1}$ ion.

In Fig. 4 we show the populations of the Ne$^{+2}$ and Ne$^{+2}(1s^{-1}-3p)$ excited states as a function of time for a field of 
peak intensity $1.0 \times 10^{17}$ W/cm$^2$. We have considered two different frequencies as in the 
case of Fig. 3, which represent the on-resonance (908.06, eV-solid lines) and off-resonance (904.06, dashed-lines) conditions.
From this figure, we can see that the populations of Ne$^{+2}$ and Ne$^{+2}(1s^{-1}-3p)$ are quickly (almost) equalized, thus allowing for the 
efficient production of the Ne$^{+3}$ ion through the Auger decay of the excited Ne$^{+2}(1s^{-1}-3p)$ state with the ejection of one more 
electron ($\Gamma_{a'} = 0.156 $ eV). With off-resonance conditions we see that the same populations evolve differently with the amount of 
Ne$^{+3}$ that is produced being significantly smaller than the Ne$^{+2}$ yield for a broad range of intensities. 
Of course, even in the detuned case, when the intensity becomes higher, the 
Rabi amplitude will increase accordingly and again the quick transfer from the ground Ne$^{+2}$ to the excited Ne$^{+2}(1s^{-1}-3p)$ state 
will allow the efficient production of the Ne$^{+3}$ ion. In that case Ne$^{+3}$ yield will surpass Ne$^{+2}$ yield. By inspection of Fig. 3, the intensities
 that this overtaking of Ne$^{+3}$ takes place are beyond $3.5\times 10^{+17}$ W/cm$^2$.   


\begin{figure}[t]
\includegraphics[scale=0.3]{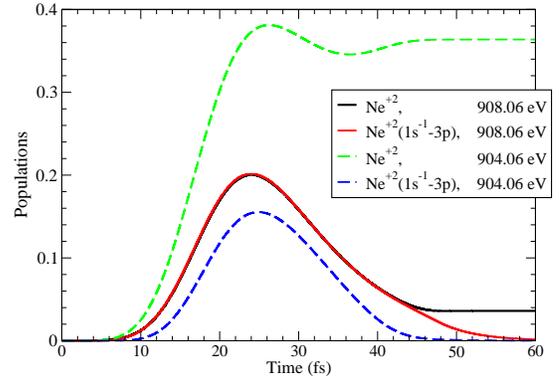}
\caption{(Color online) Populations of the Ne$^{+2}$ and Ne$^{+2}(1s^{-1}-3p)$ states as a function of time peak intensity $I_0 = 5.0 \times 10^{17}$ W/cm$^2$.
All other parameters as in Fig. 3.}
\label{fig:fig4}
\end{figure}

\section{Conclusion}
We have presented a theory of the Auger kinetic spectra and ionization yields based on the time-dependent 
density matrix theory which encapsulates all the essential dynamical parameters of the physical processes 
that are involved such as photo-ionization, photo-excitation and Auger transitions. We have examined the AES and 
the ionic yields in the case where an inner-shell photoionization takes place followed by an Auger decay of the singly charged hole-system. 
We have demonstrated the emergence of ac-Stark splitting of the Auger resonance, resulting from strong Rabi-couplings. 
In addition, ionization yields have been calculated for a range of intensities. We show how to control the branching ratios of 
various ionic species by varying dynamical parameters of the system, such as Rabi coupling and detunings. 
The theory was applied to the case of K-shell ionization of neon with the photon frequency chosen to match 
the energy differences between ionic Ne$^{+2}$ ground and excited states. 

In the present study, we have put aside the issue of a field undergoing fluctuations suitable for the description 
of more realistic situations. A more realistic approach is to assume a field with an amplitude undergoing random fluctuations 
\cite{santra:2008}. In general, the fluctuations of the field give rise to a nonzero bandwidth (beyond the Fourier bandwidth) and 
intensity fluctuations \cite{georges:1979,zoller:1982}. The main differences are that the field will excite a number of (nearby) excited states 
(see for example the most important ones in the table 1), basically those that lie within the x-ray field bandwidth 
and that the width of the Auger spectrum will be effectively increased with the x-ray bandwidth.
The development of the appropriate theory, capable to describe the field fluctuations, requires a detailed formulation which is beyond the purposes 
of the present study and it will be the subject of a future work. The essential outcome of the present work, namely the emergence of the ac-Stark splitting in the Auger 
spectra under the presence of strong ionic Rabi-couplings, will remain and the stochastic nature of the field will mainly affect its observability.

Finally, keeping the focus on the essence of the issue of experimental observability, we refrain from showing 
the influence of the volume integration in the final results, but its precise contribution needs to be evaluated 
when it comes to the actual experimental conditions.

\begin{acknowledgments}
One of us (LAAN) gratefully acknowledges discussions with Peter Lambropoulos. Useful comment from Dr. M. Meyer 
is appreciated. LAAN is mainly supported from the Science Foundation Ireland (SFI) Stokes 
Lectureship programme and by the COST Action CM0702. JC acknowledges support from SFI PI 07/IN.1/I1771 and the 
Higher Education Authority PRTLI IV and V INSPIRE programmes. TJK acknowledges the supports of an IRCSET PhD scholarship.
\end{acknowledgments}




\end{document}